\documentclass[aps,prl,twocolumn,superscriptaddress,english,floatfix]{revtex4-2}
\usepackage{graphicx}
\usepackage{float}
\usepackage{physics}
\usepackage{cancel}
\usepackage{overpic}
\usepackage{enumerate}
\usepackage{dsfont}

\usepackage[colorlinks,citecolor=blue,linkcolor=blue,urlcolor=blue]{hyperref}
\usepackage{textcomp}
\usepackage{amsmath}
\usepackage{amssymb}
\usepackage{soul}
\usepackage[normalem]{ulem}
\usepackage{lipsum}
\usepackage{comment}
\usepackage{mathrsfs}

\usepackage[english]{babel}
\usepackage{bm}
\usepackage{orcidlink}

\graphicspath{{./}{./figs/}}

\let\originalleft\left
\let\originalright\right
\renewcommand{\left}{\mathopen{}\mathclose\bgroup\originalleft}
\renewcommand{\right}{\aftergroup\egroup\originalright}

\newcommand{\vect}[1]{\boldsymbol{#1}}
\renewcommand{\vec}[1]{\vect{#1}}

\makeatletter
\newcommand*\bigcdot{{\color{gray}\mathpalette\bigcdot@{1.}}}
\newcommand*\bigcdot@[2]{\mathbin{\vcenter{\hbox{\scalebox{#2}{$\m@th#1\bullet$}}}}}
\makeatother

\def\rmi{{\rm {i}}}
\renewcommand{\d}{{\rm {d}}}
\def\dt{{\rm {d}}t}

\newcommand{\diag}{\mathrm{diag}}

\newcommand{\idhat}{\hat{\mathds{1}}}

\newcommand{\expect}[1]{\mathbb{E}\left[ #1\right]}

\newcommand{\rhohat}{\hat{\rho}}

\newcommand{\thetavec}{\vec{\theta}}

\newcommand{\schr}{Schr\"{o}dinger}

\newcommand{\aaa}{\hat{a}}
\newcommand{\daaa}{\hat{a}^\dagger}
\newcommand{\daaas}{\hat{a}^{\dagger 2}}
\newcommand{\aaas}{\aaa^{2}}

\newcommand{\Dhat}{\hat{D}}

\newcommand{\gcal}{\mathcal{G}}

\newcommand{\hh}{\hat{H}}

\newcommand{\Hhat}{\hat{H}}

\newcommand{\Lhat}{\hat{L}}
\newcommand{\dLhat}{\Lhat^\dagger}

\newcommand{\mcal}{\mathcal{M}}
\newcommand{\Mhat}{\hat{M}}

\newcommand{\ncal}{\mathcal{N}}

\newcommand{\nhat}{{\hat{n}}}

\newcommand{\Ohat}{\hat{O}}

\newcommand{\phat}{\hat{p}}
\newcommand{\pvec}{\vec{p}}
\newcommand{\pcal}{\mathcal{P}}

\newcommand{\qvec}{\vec{q}}

\newcommand{\qcal}{\mathcal{Q}}

\newcommand{\vvec}{\vec{v}}
\newcommand{\Vvec}{\vec{V}}

\newcommand{\xhat}{\hat{x}}

\newcommand{\xvec}{\vec{x}}

\newcommand{\Yvec}{\vec{Y}}

\newcommand{\zcal}{\mathcal{Z}}

\newcommand{\bea}{\begin{equation}\begin{aligned}}
		\newcommand{\eea}{\end{aligned}\end{equation}}
\newcommand{\be}{\begin{equation}}
	\newcommand{\ee}{\end{equation}}

\renewcommand{\section}[1]{%
  \noindent\textit{#1.---}
}

\begin{document}

\title{Variational Multi-Gaussian Quantum Trajectories}

\author{Zejian Li~\orcidlink{0000-0002-5652-7034}}
\affiliation{Universit\'{e} Paris Cit\'e, CNRS, Mat\'{e}riaux et Ph\'{e}nom\`{e}nes Quantiques, 75013 Paris, France}
\affiliation{Center for Theoretical Physics, Polish Academy of Sciences, Aleja Lotników 32/46, 02-668 Warsaw, Poland}

\author{Jacopo Tosca~\orcidlink{0009-0002-5165-7937}}
\affiliation{Universit\'{e} Paris Cit\'e, CNRS, Mat\'{e}riaux et Ph\'{e}nom\`{e}nes Quantiques, 75013 Paris, France}

\begin{abstract}
    We propose a variational framework for the efficient simulation of stochastic quantum dynamics in interacting bosonic systems. Our method generalizes the time-dependent variational principle to accommodate quantum state diffusion processes and is tailored for describing quantum trajectories unraveled from a master equation. We adopt a wavefunction ansatz composed of a coherent superposition of Gaussian wavepackets, which allows the fully analytical assembly of the variational equations of motion while capturing both the quantum trajectory dynamics and the Lindblad evolution beyond semiclassical approximations. The method is carefully benchmarked on a Bose-Hubbard dimer, and applied to extended Bose-Hubbard lattices in both 1D and 2D. In particular, we show the emergence of a symmetry-breaking phase transition in 2D lattices, which is absent in 1D chains.
\end{abstract}

\maketitle



\section{Introduction}
Open quantum many-body systems provide a versatile setting for exploring nonequilibrium phases and critical phenomena~\cite{fazioManybodyOpenQuantum2025,breuer2002theory}, yet their theoretical investigation is in general a formidable challenge. In interacting bosonic systems~\cite{carusottoQuantumFluidsLight2013,carusottoPhotonicMaterialsCircuit2020}, the unbounded local Hilbert space makes exact simulations rapidly intractable with increasing occupation and system size, necessitating effective descriptions. Semiclassical approaches, such as the truncated Wigner approximation~\cite{Vogel1989, POLKOVNIKOV20101790} (TWA) and Gaussian approximations~\cite{verstraelen2018gaussian,verstraelenGaussianTrajectoryApproach2020a}, provide an efficient description, yet become insufficient when interactions and non-Gaussian correlations become important.

An alternative description of open-system dynamics is offered by \textit{quantum trajectories}~\cite{Dalibard1992,Wiseman,zoller1997quantumnoise,daleyQuantumTrajectoriesOpen2014a}, which additionally provide access to the stochastic evolution of the quantum state conditioned on continuous monitoring on the wavefunction level. Their exact simulation nevertheless suffers from the same many-body complexity. Inspired by recent success in variational multi-configurational phase-space methods on the Lindblad level~\cite{toscaEfficientVariationalDynamics2025,toscaVariationalDynamicsOpen2026,toscaQuantumClassicalPotts2026}, we introduce in this work a stochastic variational multi-Gaussian (sVMG) framework for efficiently simulating stochastic quantum trajectories. We formulate a stochastic time-dependent variational principle (sTDVP) and combine it with a wavefunction ansatz consisting of coherent superpositions of Gaussian wavepackets. The resulting variational equations can be assembled analytically, enabling efficient numerical simulations, while the number of Gaussian components systematically controls the non-Gaussianity of the wavefunction.
We carefully benchmark the method on a driven-dissipative Bose-Hubbard dimer~\cite{casteelsQuantumEntanglementSpatialsymmetrybreaking2017,delmonteMeasurementinducedPhaseTransitions2025}, where sVMG reproduces exact individual trajectories and entanglement dynamics in regimes where semiclassical approximations fail. We further apply the framework to one-dimensional (1D) and 2D Bose-Hubbard lattices, showing the emergence of a symmetry-breaking transition in 2D which is suppressed in 1D.

\begin{figure*}
    \centering
    \includegraphics[width=\linewidth]{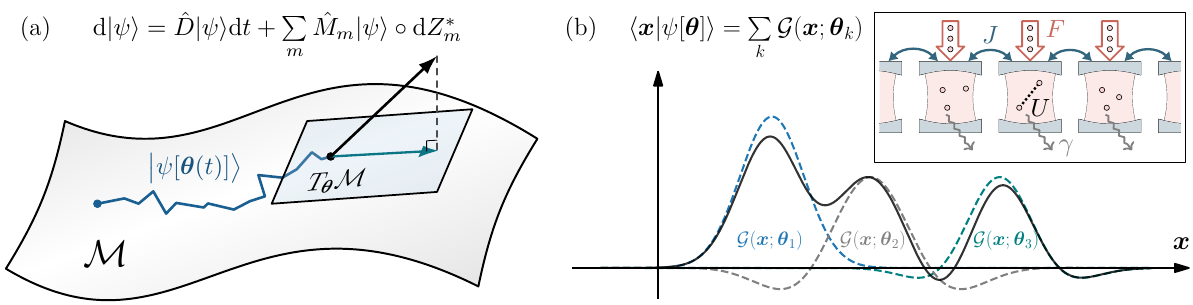}
    \caption{
    (a)~Schematic representation of the sTDVP dynamics of a variational quantum trajectory $\ket{\psi[\thetavec(t)]}$ on the variational manifold $\mcal$. At each instant $t$, the exact increment $\d\ket{\psi}$ given by the Stratonovich stochastic \schr{} equation~\eqref{eq:sse-strat} is projected onto the tangent space $T_{\thetavec}\mcal$, yielding the sTDVP equation~\eqref{eq:stdvp-eom} for $\thetavec$.    
    (b) Variational multi-Gaussian ansatz for a Bose-Hubbard model defined in~\eqref{eq:bh-ham} (sketched in the inset). The wavefunction ansatz $\braket{\xvec}{\psi[\thetavec]}$ is a coherent superposition of Gaussian wavepackets $\gcal(\xvec;\thetavec_k)$, parametrized by variational parameters $\thetavec=(\thetavec_1,\cdots,\thetavec_{N_C})$. }
    \label{fig:stdvp}
\end{figure*}

\section{Stochastic time-dependent variational principle}
\label{sec:stdvp}
We consider an open quantum system, where the dynamics of the system's density matrix $\rhohat$ is governed by the Lindblad master equation~\cite{breuer2002theory}, which reads ($\hbar = 1$)
\begin{equation}
    \label{eq:lindblad}
    \frac{\d}{\d t}\rhohat = -\rmi[\hh,\rhohat] + \sum_{m}\left( \Lhat_m\rhohat\dLhat_m - \frac{1}{2}\{ \dLhat_m\Lhat_m,\rhohat\} \right),
\end{equation}
where $\Hhat$ is the Hamiltonian, and $\Lhat_m$ is the Lindblad jump operator accounting for dissipation.

The Lindblad dynamics~\eqref{eq:lindblad} can be \textit{unraveled} into a set of \textit{quantum trajectories}. Physically, this corresponds to the stochastic evolution of a pure state $\ket{\psi}$ conditioned on the continuous measurement of the output field. In this work, we adopt the \textit{heterodyne} unraveling described by a quantum state diffusion~\cite{Wiseman}. This is commonly written as the It\^{o} stochastic \schr{} equation (SSE) \cite{Wiseman,zoller1997quantumnoise},
\begin{equation}
    \label{eq:sse-ito}
    \d\ket{\psi} = \hat{D}_{\mathrm{It\hat{o}}}\ket{\psi}\dt + \sum_m \hat{M}_m\ket{\psi}\,\d Z^*_{m}\,,
\end{equation}
where the drift and noise operators are defined as
\bea\label{eq:sse-drift-ito}
    \hat{D}_\mathrm{It\hat{o}} &= -\rmi\hh - \frac{1}{2}\sum_m\left( \dLhat_m\Lhat_m - 2\langle\dLhat_m\rangle\Lhat_m + \lvert\langle\Lhat_m\rangle\rvert^2 \right), \\
    \hat{M}_m &= \Lhat_m - \langle\Lhat_m\rangle\,,
\eea
and the noise $\d Z_m$ is a complex Wiener process satisfying $\d Z^*_{m}\d Z_{n} = \delta_{mn}\,\dt\,,\quad \d Z_{m}\d Z_{n} = 0$.
Here, $\langle\Lhat\rangle \equiv \bra{\psi}\Lhat\ket{\psi}$ is the \textit{single-trajectory} expectation value, whose dependence on $\ket{\psi}$ makes the dynamics nonlinear, accounting for the measurement back action. 
The Lindblad dynamics~\eqref{eq:lindblad} is recovered upon averaging over all possible trajectories, i.e., $\rhohat(t) = \expect{\ket{\psi(t)}\!\bra{\psi(t)}}$.

To efficiently solve the trajectory dynamics~\eqref{eq:sse-ito} for a many-body wavefunction $\ket{\psi}$, we adopt a variational approach. Let us consider a variational ansatz $\ket{\psi[\thetavec]}$ parametrized by $N_p$ real parameters $\thetavec=(\theta_1, \cdots, \theta_{N_p})$. The dynamics can then be constrained to the variational manifold $\mcal = \left\{ \ket{\psi[\thetavec]},\ \thetavec\in\mathbb{R}^{N_p}\right\}$ by projecting the exact increment $\d\ket{\psi}$ at every instant $t$ onto the tangent space $T_{\thetavec}\mcal = \mathrm{span}\{\ket{v_i}\}$ spanned by the tangent vectors $\ket{v_i} \equiv \partial_{\theta_i}\ket{\psi[\thetavec]}$ [Fig.~\ref{fig:stdvp}(a)]~\cite{yuanTheoryVariationalQuantum2019a}.
This projection, however, relies on the chain rule
\begin{equation}
    \label{eq:chain-rule}
    \d\ket{\psi[\thetavec]} = \sum_i \partial_{\theta_i}\ket{\psi[\thetavec]}\,\d\theta_i = \sum_i \ket{v_i}\,\d\theta_i\,,
\end{equation}
which is not valid in It\^{o} calculus~\cite{kloeden1992stochastic}. The ordinary chain rule is instead preserved by the Stratonovich convention when we convert the It\^{o} SSE~\eqref{eq:sse-ito} into the following equivalent Stratonovich SSE (cf. Appendix A),
\bea
    \label{eq:sse-strat}
    \d\ket{\psi} = \hat{D}\ket{\psi}\dt + \sum_m \hat{M}_m\ket{\psi}\circ\d Z^*_{m}\,,
\eea
where $\circ$ denotes the Stratonovich product and the drift
\begin{equation}
    \label{eq:drift-strat}
    \hat{D} = \hat{D}_\mathrm{It\hat{o}} + \frac{1}{2}\sum_m\left( \langle\dLhat_m\Lhat_m\rangle - \lvert\langle\Lhat_m\rangle\rvert^2 \right)
\end{equation}
is redefined by 
absorbing a correction term. Inserting the chain rule~\eqref{eq:chain-rule} into the Stratonovich equation~\eqref{eq:sse-strat} and projecting both sides onto the tangent vector $\bra{v_j}$ yields~\footnote{With real variational parameters, the tangent space $T_{\thetavec}\mcal$ is regarded as a real vector space endowed with the inner product $(\psi_1,\psi_2)=\Re\braket{\psi_1}{\psi_2}$. The corresponding orthogonal projection is then equivalent to the McLachlan variational principle~\cite{yuanTheoryVariationalQuantum2019a}, which minimizes the Hilbert-space distance between the exact and variational increments.}
\begin{equation}
    \label{eq:galerkin}
    \Re\sum_i S_{ji}\,\d\theta_i = \Re\Big(\bra{v_j}\hat{D}\ket{\psi}\dt + \sum_m \bra{v_j}\hat{M}_m\ket{\psi}\circ\d Z^*_{m}\Big)\,,
\end{equation}
where $S_{ji} \equiv \braket{v_j}{v_i}$ is the \textit{quantum geometric tensor}. Denoting the drift and noise forces $V_{j} = \bra{v_j}\hat{D}\ket{\psi}$, $Y^{(m)}_{j} = \bra{v_j}\hat{M}_m\ket{\psi}$,
the variational dynamics can be formally written as
\begin{equation}
    \label{eq:stdvp-eom}
    \d\thetavec = \Re\left(S^{-1}\right)\Re\left( \vec{V}\dt + \sum_m \vec{Y}^{(m)}\circ\d Z^*_{m} \right),
\end{equation}
which constitutes the stochastic time-dependent variational principle (sTDVP).

\begin{figure*}[t]
    \centering
    \includegraphics[width=\linewidth]{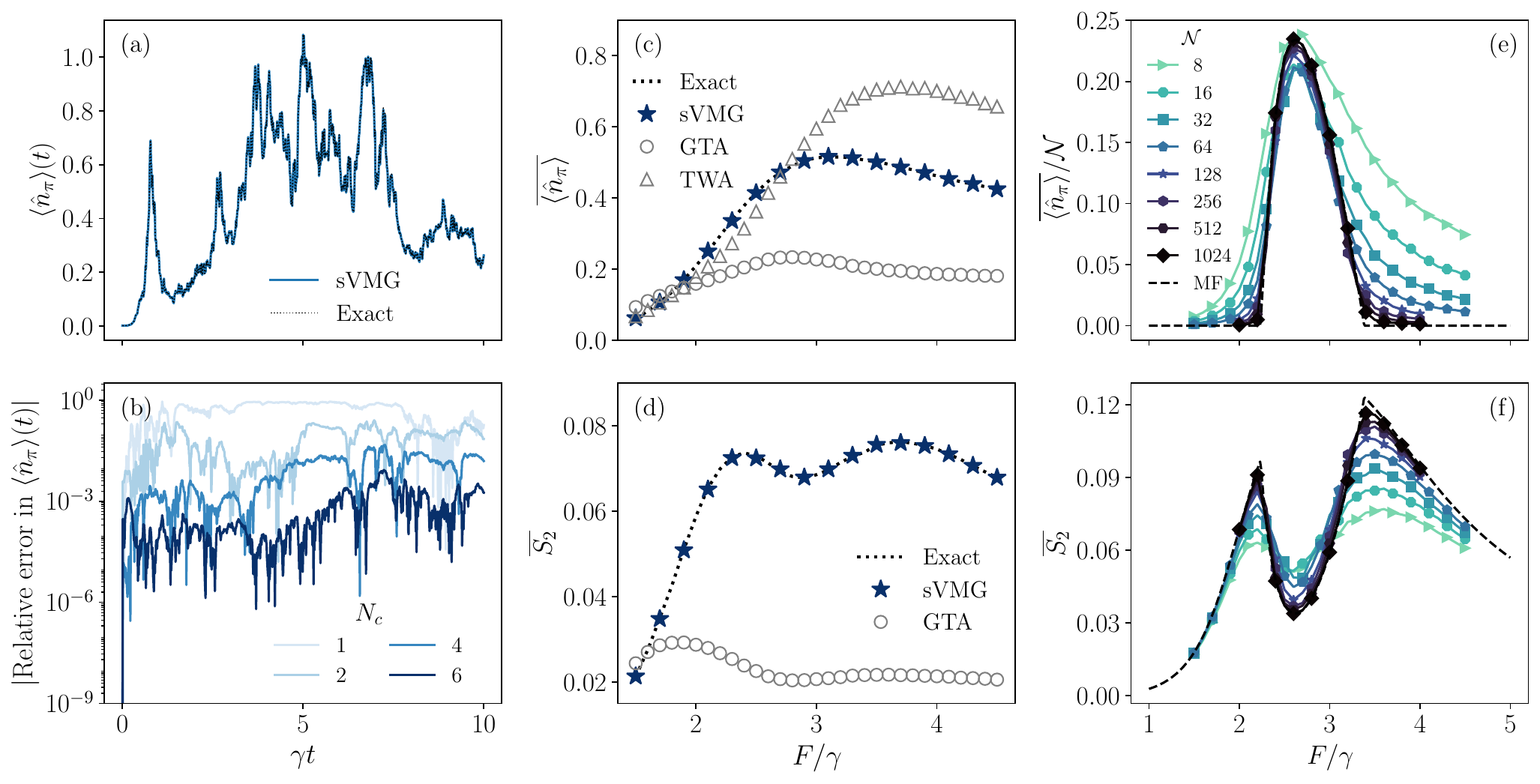}
    \caption{Results on the Bose-Hubbard dimer.
    (a) Population of the antibonding mode $\langle\nhat_\pi\rangle$ along a single trajectory with $F=3\gamma$, obtained with the stochastic variational multi-gaussian (sVMG) approach using $N_c=4$ Gaussian components, together with the exact solution. (b) Relative error of the sVMG result along the trajectory in (a), defined as $|\langle\nhat\rangle^{\mathrm{sVMG}}-\langle\nhat\rangle^{\mathrm{Exact}}|/|\langle\nhat\rangle^{\mathrm{Exact}}|$, for different values of $N_c$. (c)-(d) Steady-state values of the trajectory-averaged antibonding mode population $\overline{\langle\nhat\rangle}$ and the bipartite R\'{e}nyi-2 entanglement entropy $\overline{S_2}$ as a function of the drive $F$, obtained with sVMG ($N_c=6$), the truncated Wigner approximation (TWA, not applicable to $\overline{S_2}$), the Gaussian trajectory approximation (GTA), and the exact solution. (e)-(f) sVMG steady-state results for the rescaled Bose-Hubbard dimer with $F\to F\sqrt{\ncal}, U\to U/\ncal$, of the normalized antibonding mode population $\overline{\langle\nhat_\pi\rangle}/\ncal$ and the R\'{e}nyi-2 entanglement entropy $\overline{S_2}$, as a function of the drive $F$, for different values of $\ncal$ shown together with the mean-field (MF) solution. Parameters: $U=2\gamma$, $J=-2.5\gamma$, $\Delta=-1.4\gamma$.}
    \label{fig:svmg-bhd-bench}
\end{figure*}

\section{Multi-Gaussian wavefunction ansatz for bosonic systems}
\label{sec:ansatz}
To demonstrate the sTDVP protocol, we consider a driven-dissipative Bose-Hubbard lattice~\cite{leboiteSteadyStatePhasesTunnelingInduced2013} with $N$ sites, as sketched in the inset of Fig.~\ref{fig:stdvp} (b). The Hamiltonian is defined as~\footnote{In the frame rotating at the drive frequency with $\hbar=1$.},
\bea\label{eq:bh-ham}
    \Hhat &= -\Delta\sum_i \daaa_i\aaa_i + \frac{U}{2}\sum_i\daaas_i\aaas_i+F(\daaa_i+\aaa_i)\\
    &~~~ -J\sum_{\langle i,j\rangle}(\daaa_i\aaa_j+\daaa_j\aaa_i)\,,
\eea
where $\aaa_i$ is the annihilation operator for site $i$ with standard commutation relations $[\aaa_i,\daaa_j]=\delta_{ij}$, $\Delta$ is the detuning between the drive and the bare mode frequencies, $U$ is the on-site self-interaction strength, $F$ is the amplitude of a uniform coherent drive, $J$ is the boson hopping amplitude along lattice edges $\langle i,j\rangle$. Each site is subjected to single-body loss $\Lhat_i=\sqrt{\gamma}\aaa_i$ with rate $\gamma$. 

The problem can be equivalently formulated in the position representation exploiting the canonical mapping,
\bea\label{eq:canonical-mapping}
    \ket{\psi}&\mapsto \braket{\xvec}{\psi}\equiv \psi(\xvec)\,,\\
    \aaa_i^{(\dagger)} =(\xhat_i\pm\rmi\phat_i)/\sqrt{2} &\mapsto \left(x_i\pm{\partial_{x_i}}\right)/\sqrt{2}\,,
\eea
with the position coordinate $\xvec=(x_1,\cdots, x_N)\in\mathbb{R}^N$. 
To solve the stochastic dynamics~\eqref{eq:sse-strat}, we adopt a variational wavefunction ansatz consisting of a coherent superposition of $N_c$ Gaussian wavepackets~\cite{worthFullQuantumMechanical2003,richingsQuantumDynamicsSimulations2015,lasserComputingQuantumDynamics2020}:
\bea\label{eq:psi-vmg}
    \braket{\xvec}{\psi[\thetavec]} &= \sum_{k=1}^{N_c} \gcal(\xvec;\thetavec_k)\,, \quad \thetavec=(\thetavec_1,\cdots, \thetavec_{N_c})\,,
\eea
where each component reads
\bea\label{eq:Gkx}
    \gcal(\xvec;\thetavec_k)&= C_k \nu_k\times\\
    &\hspace{-0.8cm}\exp\big\{-(\xvec-\qvec_k)^TA_k(\xvec-\qvec_k)+\rmi \pvec_k^T(\xvec-\qvec_k) \big\}\,,
\eea
with $\nu_k=\pi^{-N/4}\det(A_k+A^*_k)^{1/4}$ the local normalization factor, $\qvec_k,\pvec_k\in\mathbb{R}^N$ the mean
position and momentum, $A_k$ a complex symmetric $N\times N$ matrix  setting the covariance, and $C_k\in\mathbb{C}$ is the weight respecting the overall normalization $\braket{\psi[\thetavec]}{\psi[\thetavec]}=1$.  For large systems, one can further compress the ansatz via a ``low-rank'' decomposition of the covariances, $A_k=\diag(\lambda^{(k)}_1,\cdots,\lambda^{(k)}_N)+\Lambda_k\Lambda_k^T$, with $\lambda^{(k)}_i\in\mathbb{C}$ and $\Lambda_k\in \mathbb{C}^{N\times K}$ with $1\le K\le N$ controlling the rank of the off-diagonal part of $A_k$, giving a total of $N_p=2N_c(1+2N+NK)$ real parameters~\footnote{For a complex parametrization, we split the real and imaginary parts and work with the effective real parameters.}.

Importantly, this ansatz allows the sTDVP equations~\eqref{eq:stdvp-eom} to be assembled \textit{analytically}: the quantum geometric tensor $S$ and the force vectors $\vec{V}$, $\vec{Y}^{(m)}$ in Eq.~\eqref{eq:stdvp-eom} fully rely on integrating polynomial-weighted Gaussian functions (cf. Appendix B), a class of analytically integrable functions thanks to Wick's theorem. The resulting equation~\eqref{eq:stdvp-eom} for the parameters $\thetavec$ can be solved efficiently as a standard Stratonovich stochastic differential equation~\footnote{Due to the overcompleteness of the Gaussian basis, we employ a standard Tikhonov regularization for numerical stability, which amounts to adding a small diagonal shift $\epsilon$ (typically $\in [10^{-12},10^{-6}]$) to the quantum geometric tensor $S$ before inverting it.}. We adopt in this work the standard Euler-Heun method~\cite{ruemelinNumericalTreatmentStochastic1982}, which discretizes the time into fixed steps $\delta t$ to integrate Eq.~\eqref{eq:stdvp-eom} numerically.

\begin{figure*}[t]

    \centering
    \includegraphics[width=\linewidth]{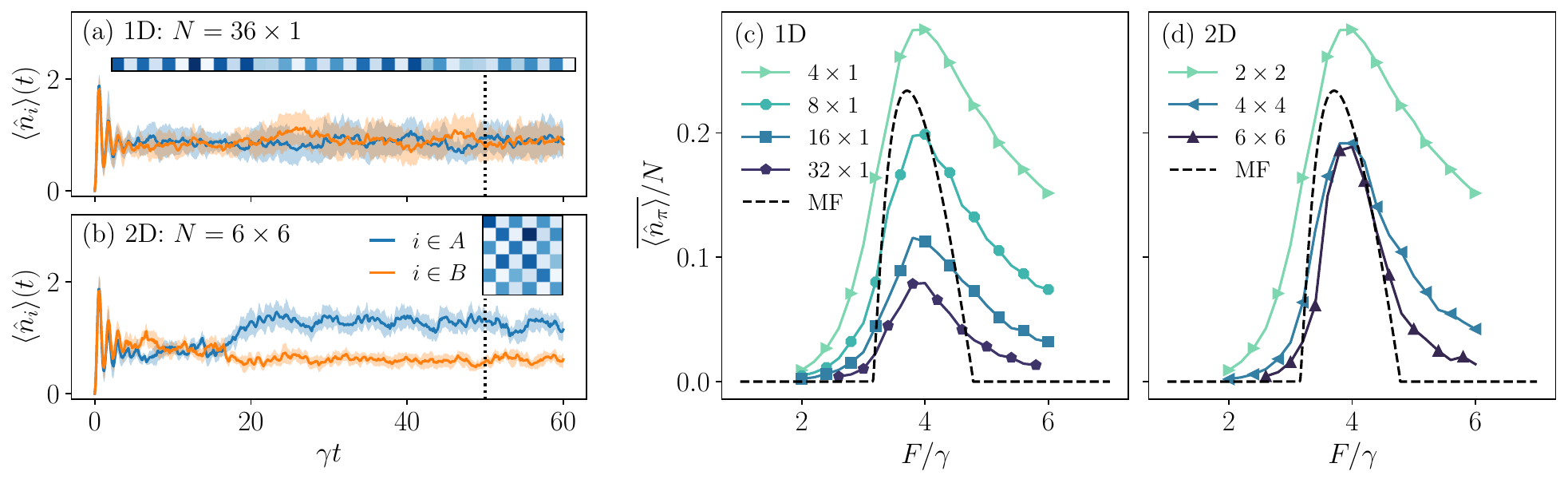}
    \caption{Results on periodic Bose-Hubbard lattices. (a)-(b) Example single trajectories (at $F=4\gamma$) of a 1D chain with 36 sites and of a 2D $6\times 6$ lattice, represented as the average population in two checkerboard sublattices ($A$ and $B$) with the shaded area indicating the standard deviation within each sublattice. Inset: snapshot of the population $\langle\nhat_i\rangle$ distribution over the lattice at $\gamma t=50$ (with color intensity proportional to $\langle\nhat_i\rangle$). (c)-(d) Steady-state normalized $\pi-$mode population $\langle\nhat_\pi\rangle/N$ as a function of the drive in 1D and 2D, for different lattice sizes. Dashed line indicates the mean-field (MF) result for the dimer model. Parameters: $U=\gamma, zJ=-2.5\gamma, \Delta=-1.4\gamma$. }
    \label{fig:bh-1d-vs-2d}    
\end{figure*}

\section{Bose-Hubbard dimer}
As a first demonstration of the stochastic variational multi-Gaussian (sVMG) framework, we consider the special case of the Bose-Hubbard dimer with $N=2$ sites. When the boson population is low, the stochastic dynamics~\eqref{eq:sse-strat} can be integrated exactly (by truncating the Fock space), providing a benchmark for sVMG. We consider the parameters $U=2\gamma, J=-2.5\gamma$ and $\Delta=-1.4\gamma$, a regime where the mean-field theory predicts spontaneous breaking of the dimer's $\mathbb{Z}_2$ spatial symmetry~\cite{casteelsQuantumEntanglementSpatialsymmetrybreaking2017}. For finite systems, this signature can be probed via the antibonding mode population~\footnote{Due to the spatial symmetry of the Liouvillian, the (unique) steady state will have uniform population even in the mean-field symmetry-broken regime~\cite{mingantiSpectralTheoryLiouvillians2018a}. Instead, $\langle\nhat_\pi\rangle$ can remain nonzero both on the Lindblad level and along trajectories.} $\langle\nhat_\pi\rangle$ with $\nhat_\pi=\daaa_\pi\aaa_\pi$ with $\aaa_\pi\equiv(\aaa_1-\aaa_2)/\sqrt{2}$. Figure~\ref{fig:svmg-bhd-bench} (a) shows the single-trajectory expectation value $\langle\nhat_\pi\rangle$ obtained with sVMG with $N_c=4$ Gaussian components at $F=3\gamma$, which faithfully reproduces the exact solution obtained with the same time step ($\gamma\delta t=10^{-4}$) and noise realization. Panel (b) shows the relative error in $\langle\nhat_\pi\rangle$ along a single trajectory, which decreases with $N_c$, showing the controlled convergence of the sVMG method towards the exact dynamics. Panels (c) and (d) show the benchmark of trajectory-average results for the observable $\langle\nhat_\pi\rangle$ and also the Rényi-2 entanglement entropy $\overline{S_2}$ [defined as $S_2=-\ln \Tr(\rhohat_1^2)$ with $\rhohat_1=\Tr_2(\ketbra{\psi})$], where the sVMG ($N_c$=6) reproduces the exact solution for a wide range of the drive $F$. For comparison, we show the results obtained with the Gaussian trajectory approximation (equivalent to sVMG with $N_c=1$) and the truncated Wigner approximation (applicable only to the linear observable $\langle\nhat_\pi\rangle$), both of which fail to capture the exact dynamics. This benchmark also highlights the ability of sVMG to evaluate \textit{nonlinear} state functionals (such as entanglement) along single trajectories, inaccessible to methods operating on the Lindblad level. 

The Bose-Hubbard dimer admits a well-defined thermodynamic limit under the scaling of parameters $F\to F\sqrt{\ncal}, U\to U/\ncal$~\cite{casteelsQuantumEntanglementSpatialsymmetrybreaking2017,casteelsCriticalDynamicalProperties2017}, where the large $\ncal$ limit corresponds to having a large photon population in each mode, a regime beyond the reach of brute-force exact simulation yet perfectly accessible via the sVMG approach.  Fig.~\ref{fig:svmg-bhd-bench} (e) and (f) present the sVMG results for the steady-state $\langle\nhat_\pi\rangle$ and the trajectory-averaged entanglement $\overline{S_2}$ of the rescaled Bose-Hubbard dimer. As $\ncal$ increases, we observe the finite-$\ncal$ results approach the mean-field solution~\footnote{The ``mean-field'' trajectory entanglement is obtained via a semiclassical theory recently developed by some of us as detailed in~\cite{liEmergentDeterministicEntanglement2025}.}: a symmetry-broken phase with $\langle\nhat_\pi\rangle\neq 0$ emerges between two critical values of $F$, accompanied by an entanglement transition occurring at the same critical points. This confirms the results of previous studies obtained with the semiclassical GTA approach~\cite{delmonteMeasurementinducedPhaseTransitions2025}. We also report that a maximum of $N_c=4$ components is sufficient for obtaining convergent results for all values of $\ncal$, which is consistent with the infinite-range nature of the model, where one generally expects the non-Gaussianity to decrease as $\ncal$ increases~\cite{liEmergentDeterministicEntanglement2025}.

\section{Bose-Hubbard lattices}
Let us now consider periodic Bose-Hubbard lattices in 1D and 2D, and seek the analogue of the dimer's symmetry-breaking transition, corresponding to the formation of checkerboard patterns on the lattice. We consider the parameters $U=\gamma, zJ=-2.5\gamma$ and $\Delta=-1.4\gamma$ ($z$ is the lattice coordination number). Figures~\ref{fig:bh-1d-vs-2d} (a) and (b) show the distribution of the populations $\langle\nhat_i\rangle$ in the two checkerboard sublattices $A$ and $B$ [defined as $A,B=\{i~|~(-1)^{i_x+i_y}=\pm 1\}$, where $(i_x,i_y)$ are integer coordinates of the $i-$th site in the lattice] along a single trajectory for a $1\times36$ 1D chain and a $6\times6$ 2D lattice, together with a snapshot of the population profile over the lattice at $\gamma t=50$. While the 2D population displays a checkerboard pattern with a clear separation between the two sublattices, this is not the case in 1D: we observe local checkerboard-like regions with opposite phases separated by domain walls, resulting in a lower difference in the average populations over the two sublattices. This suggests that the symmetry-broken phase might survive in 2D, yet can be washed out in 1D due to enhanced fluctuations. 
To verify this, we consider again the steady-state order parameter $\langle\nhat_\pi\rangle$, the population of the $\pi$-momentum mode, where the $\pi-$mode operator is now $\aaa_\pi=\sum_{i_x, i_y}(-1)^{i_x+i_y}\aaa_i/\sqrt{N}$. Figures~\ref{fig:bh-1d-vs-2d} (c) and (d) show the normalized quantity $\langle\nhat_\pi\rangle/N$ as a function of the drive $F$ in 1D and 2D, for different lattice sizes $N$. A maximum of $N_c=4$ and $K=2$ is used in the low-rank ansatz for convergence of the results (cf. Appendix C). As $N$ increases, the 1D results decrease towards zero, suggesting failure to establish a long-range $\pi-$mode order, while the 2D normalized $\pi-$mode population remains finite and converges to a dome shape qualitatively similar to the dimer's mean-field solution, suggesting the emergence of a symmetry-broken phase in the thermodynamic limit, consistent with our expectations from single-trajectory observations. 

\section{Conclusion}
We have introduced a stochastic variational multi-Gaussian (sVMG) framework for simulating quantum trajectories in interacting bosonic systems, capable of capturing quantum correlations beyond conventional semiclassical approaches, and the fully analytical structure of the variational equations allows efficient simulations of many-body systems. Applying this approach to a driven-dissipative Bose-Hubbard model, our results suggest that the 2D lattice exhibits a symmetry-breaking transition similar to the one known in the Bose-Hubbard dimer, while the transition does not survive in 1D. We expect this dimension-dependent criticality to be readily observable in experiments on photonic platforms~\cite{Zejian}. Our framework can be readily generalized to incorporate spin systems~\cite{toscaVariationalDynamicsOpen2026,liMonitoredLongrangeInteracting2025a,liGeneralizedStochasticSpinwave2026}, and applies to both the Lindblad level and measurement-induced dynamics along trajectories, opening up a new avenue for exploring non-equilibrium quantum dynamics.

\bibliography{biblo}

\begin{thebibliography}{37}%
\makeatletter
\providecommand \@ifxundefined [1]{%
 \@ifx{#1\undefined}
}%
\providecommand \@ifnum [1]{%
 \ifnum #1\expandafter \@firstoftwo
 \else \expandafter \@secondoftwo
 \fi
}%
\providecommand \@ifx [1]{%
 \ifx #1\expandafter \@firstoftwo
 \else \expandafter \@secondoftwo
 \fi
}%
\providecommand \natexlab [1]{#1}%
\providecommand \enquote  [1]{``#1''}%
\providecommand \bibnamefont  [1]{#1}%
\providecommand \bibfnamefont [1]{#1}%
\providecommand \citenamefont [1]{#1}%
\providecommand \href@noop [0]{\@secondoftwo}%
\providecommand \href [0]{\begingroup \@sanitize@url \@href}%
\providecommand \@href[1]{\@@startlink{#1}\@@href}%
\providecommand \@@href[1]{\endgroup#1\@@endlink}%
\providecommand \@sanitize@url [0]{\catcode `\\12\catcode `\$12\catcode `\&12\catcode `\#12\catcode `\^12\catcode `\_12\catcode `\%12\relax}%
\providecommand \@@startlink[1]{}%
\providecommand \@@endlink[0]{}%
\providecommand \url  [0]{\begingroup\@sanitize@url \@url }%
\providecommand \@url [1]{\endgroup\@href {#1}{\urlprefix }}%
\providecommand \urlprefix  [0]{URL }%
\providecommand \Eprint [0]{\href }%
\providecommand \doibase [0]{https://doi.org/}%
\providecommand \selectlanguage [0]{\@gobble}%
\providecommand \bibinfo  [0]{\@secondoftwo}%
\providecommand \bibfield  [0]{\@secondoftwo}%
\providecommand \translation [1]{[#1]}%
\providecommand \BibitemOpen [0]{}%
\providecommand \bibitemStop [0]{}%
\providecommand \bibitemNoStop [0]{.\EOS\space}%
\providecommand \EOS [0]{\spacefactor3000\relax}%
\providecommand \BibitemShut  [1]{\csname bibitem#1\endcsname}%
\let\auto@bib@innerbib\@empty
\bibitem [{\citenamefont {Fazio}\ \emph {et~al.}(2025)\citenamefont {Fazio}, \citenamefont {Keeling}, \citenamefont {Mazza},\ and\ \citenamefont {Schir{\`o}}}]{fazioManybodyOpenQuantum2025}%
  \BibitemOpen
  \bibfield  {author} {\bibinfo {author} {\bibfnamefont {R.}~\bibnamefont {Fazio}}, \bibinfo {author} {\bibfnamefont {J.}~\bibnamefont {Keeling}}, \bibinfo {author} {\bibfnamefont {L.}~\bibnamefont {Mazza}},\ and\ \bibinfo {author} {\bibfnamefont {M.}~\bibnamefont {Schir{\`o}}},\ }\bibfield  {title} {\bibinfo {title} {Many-body open quantum systems},\ }\href {https://doi.org/10.21468/SciPostPhysLectNotes.99} {\bibfield  {journal} {\bibinfo  {journal} {SciPost Physics Lecture Notes}\ ,\ \bibinfo {pages} {099}} (\bibinfo {year} {2025})}\BibitemShut {NoStop}%
\bibitem [{\citenamefont {Breuer}\ and\ \citenamefont {Petruccione}(2002)}]{breuer2002theory}%
  \BibitemOpen
  \bibfield  {author} {\bibinfo {author} {\bibfnamefont {H.-P.}\ \bibnamefont {Breuer}}\ and\ \bibinfo {author} {\bibfnamefont {F.}~\bibnamefont {Petruccione}},\ }\href@noop {} {\emph {\bibinfo {title} {The theory of open quantum systems}}}\ (\bibinfo  {publisher} {OUP Oxford},\ \bibinfo {year} {2002})\BibitemShut {NoStop}%
\bibitem [{\citenamefont {Carusotto}\ and\ \citenamefont {Ciuti}(2013)}]{carusottoQuantumFluidsLight2013}%
  \BibitemOpen
  \bibfield  {author} {\bibinfo {author} {\bibfnamefont {I.}~\bibnamefont {Carusotto}}\ and\ \bibinfo {author} {\bibfnamefont {C.}~\bibnamefont {Ciuti}},\ }\bibfield  {title} {\bibinfo {title} {Quantum fluids of light},\ }\href {https://doi.org/10.1103/RevModPhys.85.299} {\bibfield  {journal} {\bibinfo  {journal} {Reviews of Modern Physics}\ }\textbf {\bibinfo {volume} {85}},\ \bibinfo {pages} {299} (\bibinfo {year} {2013})}\BibitemShut {NoStop}%
\bibitem [{\citenamefont {Carusotto}\ \emph {et~al.}(2020)\citenamefont {Carusotto}, \citenamefont {Houck}, \citenamefont {Koll{\'a}r}, \citenamefont {Roushan}, \citenamefont {Schuster},\ and\ \citenamefont {Simon}}]{carusottoPhotonicMaterialsCircuit2020}%
  \BibitemOpen
  \bibfield  {author} {\bibinfo {author} {\bibfnamefont {I.}~\bibnamefont {Carusotto}}, \bibinfo {author} {\bibfnamefont {A.~A.}\ \bibnamefont {Houck}}, \bibinfo {author} {\bibfnamefont {A.~J.}\ \bibnamefont {Koll{\'a}r}}, \bibinfo {author} {\bibfnamefont {P.}~\bibnamefont {Roushan}}, \bibinfo {author} {\bibfnamefont {D.~I.}\ \bibnamefont {Schuster}},\ and\ \bibinfo {author} {\bibfnamefont {J.}~\bibnamefont {Simon}},\ }\bibfield  {title} {\bibinfo {title} {Photonic materials in circuit quantum electrodynamics},\ }\href {https://doi.org/10.1038/s41567-020-0815-y} {\bibfield  {journal} {\bibinfo  {journal} {Nature Physics}\ }\textbf {\bibinfo {volume} {16}},\ \bibinfo {pages} {268} (\bibinfo {year} {2020})}\BibitemShut {NoStop}%
\bibitem [{\citenamefont {Vogel}\ and\ \citenamefont {Risken}(1989)}]{Vogel1989}%
  \BibitemOpen
  \bibfield  {author} {\bibinfo {author} {\bibfnamefont {K.}~\bibnamefont {Vogel}}\ and\ \bibinfo {author} {\bibfnamefont {H.}~\bibnamefont {Risken}},\ }\bibfield  {title} {\bibinfo {title} {Quasiprobability distributions in dispersive optical bistability},\ }\href {https://doi.org/10.1103/physreva.39.4675} {\bibfield  {journal} {\bibinfo  {journal} {Phys. Rev. A}\ }\textbf {\bibinfo {volume} {39}},\ \bibinfo {pages} {4675} (\bibinfo {year} {1989})}\BibitemShut {NoStop}%
\bibitem [{\citenamefont {Polkovnikov}(2010)}]{POLKOVNIKOV20101790}%
  \BibitemOpen
  \bibfield  {author} {\bibinfo {author} {\bibfnamefont {A.}~\bibnamefont {Polkovnikov}},\ }\bibfield  {title} {\bibinfo {title} {Phase space representation of quantum dynamics},\ }\href {https://doi.org/https://doi.org/10.1016/j.aop.2010.02.006} {\bibfield  {journal} {\bibinfo  {journal} {Annals of Physics}\ }\textbf {\bibinfo {volume} {325}},\ \bibinfo {pages} {1790} (\bibinfo {year} {2010})}\BibitemShut {NoStop}%
\bibitem [{\citenamefont {Verstraelen}\ and\ \citenamefont {Wouters}(2018)}]{verstraelen2018gaussian}%
  \BibitemOpen
  \bibfield  {author} {\bibinfo {author} {\bibfnamefont {W.}~\bibnamefont {Verstraelen}}\ and\ \bibinfo {author} {\bibfnamefont {M.}~\bibnamefont {Wouters}},\ }\bibfield  {title} {\bibinfo {title} {Gaussian quantum trajectories for the variational simulation of open quantum-optical systems},\ }\href {https://doi.org/10.3390/app8091427} {\bibfield  {journal} {\bibinfo  {journal} {App. Sci.}\ }\textbf {\bibinfo {volume} {8}},\ \bibinfo {pages} {1427} (\bibinfo {year} {2018})}\BibitemShut {NoStop}%
\bibitem [{\citenamefont {Verstraelen}\ \emph {et~al.}(2020)\citenamefont {Verstraelen}, \citenamefont {Rota}, \citenamefont {Savona},\ and\ \citenamefont {Wouters}}]{verstraelenGaussianTrajectoryApproach2020a}%
  \BibitemOpen
  \bibfield  {author} {\bibinfo {author} {\bibfnamefont {W.}~\bibnamefont {Verstraelen}}, \bibinfo {author} {\bibfnamefont {R.}~\bibnamefont {Rota}}, \bibinfo {author} {\bibfnamefont {V.}~\bibnamefont {Savona}},\ and\ \bibinfo {author} {\bibfnamefont {M.}~\bibnamefont {Wouters}},\ }\bibfield  {title} {\bibinfo {title} {Gaussian trajectory approach to dissipative phase transitions: {{The}} case of quadratically driven photonic lattices},\ }\href {https://doi.org/10.1103/PhysRevRes..2.022037} {\bibfield  {journal} {\bibinfo  {journal} {Phys. Rev. Res.}\ }\textbf {\bibinfo {volume} {2}},\ \bibinfo {pages} {022037} (\bibinfo {year} {2020})}\BibitemShut {NoStop}%
\bibitem [{\citenamefont {Dalibard}\ \emph {et~al.}(1992)\citenamefont {Dalibard}, \citenamefont {Castin},\ and\ \citenamefont {M\o{}lmer}}]{Dalibard1992}%
  \BibitemOpen
  \bibfield  {author} {\bibinfo {author} {\bibfnamefont {J.}~\bibnamefont {Dalibard}}, \bibinfo {author} {\bibfnamefont {Y.}~\bibnamefont {Castin}},\ and\ \bibinfo {author} {\bibfnamefont {K.}~\bibnamefont {M\o{}lmer}},\ }\bibfield  {title} {\bibinfo {title} {Wave-function approach to dissipative processes in quantum optics},\ }\href {https://doi.org/10.1103/PhysRevLett.68.580} {\bibfield  {journal} {\bibinfo  {journal} {Phys. Rev. Lett.}\ }\textbf {\bibinfo {volume} {68}},\ \bibinfo {pages} {580} (\bibinfo {year} {1992})}\BibitemShut {NoStop}%
\bibitem [{\citenamefont {Wiseman}\ and\ \citenamefont {Milburn}(2009)}]{Wiseman}%
  \BibitemOpen
  \bibfield  {author} {\bibinfo {author} {\bibfnamefont {H.~M.}\ \bibnamefont {Wiseman}}\ and\ \bibinfo {author} {\bibfnamefont {G.~J.}\ \bibnamefont {Milburn}},\ }\href@noop {} {\emph {\bibinfo {title} {Quantum measurement and control}}}\ (\bibinfo  {publisher} {Cambridge university press},\ \bibinfo {year} {2009})\BibitemShut {NoStop}%
\bibitem [{\citenamefont {Zoller}\ and\ \citenamefont {Gardiner}(1997)}]{zoller1997quantumnoise}%
  \BibitemOpen
  \bibfield  {author} {\bibinfo {author} {\bibfnamefont {P.}~\bibnamefont {Zoller}}\ and\ \bibinfo {author} {\bibfnamefont {C.~W.}\ \bibnamefont {Gardiner}},\ }\bibfield  {title} {\bibinfo {title} {Quantum noise in quantum optics: the stochastic {S}chr\"{o}dinger equation},\ }in\ \href {https://arxiv.org/abs/quant-ph/9702030} {\emph {\bibinfo {booktitle} {Quantum Fluctuations, Les Houches Session LXIII}}},\ \bibinfo {editor} {edited by\ \bibinfo {editor} {\bibfnamefont {S.}~\bibnamefont {Reynaud}}, \bibinfo {editor} {\bibfnamefont {E.}~\bibnamefont {Giacobino}},\ and\ \bibinfo {editor} {\bibfnamefont {J.}~\bibnamefont {Zinn-Justin}}}\ (\bibinfo  {publisher} {Elsevier},\ \bibinfo {address} {Amsterdam},\ \bibinfo {year} {1997})\ \Eprint {https://arxiv.org/abs/quant-ph/9702030} {arXiv:quant-ph/9702030} \BibitemShut {NoStop}%
\bibitem [{\citenamefont {Daley}(2014)}]{daleyQuantumTrajectoriesOpen2014a}%
  \BibitemOpen
  \bibfield  {author} {\bibinfo {author} {\bibfnamefont {A.~J.}\ \bibnamefont {Daley}},\ }\bibfield  {title} {\bibinfo {title} {Quantum trajectories and open many-body quantum systems},\ }\href {https://doi.org/10.1080/00018732.2014.933502} {\bibfield  {journal} {\bibinfo  {journal} {Advances in Physics}\ }\textbf {\bibinfo {volume} {63}},\ \bibinfo {pages} {77} (\bibinfo {year} {2014})}\BibitemShut {NoStop}%
\bibitem [{\citenamefont {Tosca}\ \emph {et~al.}(2025)\citenamefont {Tosca}, \citenamefont {Carnazza}, \citenamefont {Giacomelli},\ and\ \citenamefont {Ciuti}}]{toscaEfficientVariationalDynamics2025}%
  \BibitemOpen
  \bibfield  {author} {\bibinfo {author} {\bibfnamefont {J.}~\bibnamefont {Tosca}}, \bibinfo {author} {\bibfnamefont {F.}~\bibnamefont {Carnazza}}, \bibinfo {author} {\bibfnamefont {L.}~\bibnamefont {Giacomelli}},\ and\ \bibinfo {author} {\bibfnamefont {C.}~\bibnamefont {Ciuti}},\ }\href {https://doi.org/10.48550/arXiv.2507.14076} {\bibinfo {title} {Efficient {{Variational Dynamics}} of {{Open Quantum Bosonic Systems}} via {{Automatic Differentiation}}}} (\bibinfo {year} {2025}),\ \bibinfo {note} {accepted for publication in Phys. Rev. X},\ \Eprint {https://arxiv.org/abs/2507.14076} {arXiv:2507.14076} \BibitemShut {NoStop}%
\bibitem [{\citenamefont {Tosca}\ \emph {et~al.}(2026{\natexlab{a}})\citenamefont {Tosca}, \citenamefont {Li}, \citenamefont {Carnazza},\ and\ \citenamefont {Ciuti}}]{toscaVariationalDynamicsOpen2026}%
  \BibitemOpen
  \bibfield  {author} {\bibinfo {author} {\bibfnamefont {J.}~\bibnamefont {Tosca}}, \bibinfo {author} {\bibfnamefont {Z.}~\bibnamefont {Li}}, \bibinfo {author} {\bibfnamefont {F.}~\bibnamefont {Carnazza}},\ and\ \bibinfo {author} {\bibfnamefont {C.}~\bibnamefont {Ciuti}},\ }\href {https://doi.org/10.48550/arXiv.2604.01165} {\bibinfo {title} {Variational {{Dynamics}} of {{Open Quantum Spin Systems}} in {{Phase Space}}}} (\bibinfo {year} {2026}{\natexlab{a}}),\ \Eprint {https://arxiv.org/abs/2604.01165} {arXiv:2604.01165} \BibitemShut {NoStop}%
\bibitem [{\citenamefont {Tosca}\ \emph {et~al.}(2026{\natexlab{b}})\citenamefont {Tosca}, \citenamefont {Li},\ and\ \citenamefont {Ciuti}}]{toscaQuantumClassicalPotts2026}%
  \BibitemOpen
  \bibfield  {author} {\bibinfo {author} {\bibfnamefont {J.}~\bibnamefont {Tosca}}, \bibinfo {author} {\bibfnamefont {Z.}~\bibnamefont {Li}},\ and\ \bibinfo {author} {\bibfnamefont {C.}~\bibnamefont {Ciuti}},\ }\bibfield  {title} {\bibinfo {title} {Quantum and {{Classical Potts Criticality}} in {{Driven-Dissipative Bosonic Lattices}}},\ }\href {https://arxiv.org/abs/2607.08425} {\bibfield  {journal} {\bibinfo  {journal} {arXiv:2607.08425}\ } (\bibinfo {year} {2026}{\natexlab{b}})},\ \Eprint {https://arxiv.org/abs/2607.08425} {2607.08425} \BibitemShut {NoStop}%
\bibitem [{\citenamefont {Casteels}\ and\ \citenamefont {Ciuti}(2017)}]{casteelsQuantumEntanglementSpatialsymmetrybreaking2017}%
  \BibitemOpen
  \bibfield  {author} {\bibinfo {author} {\bibfnamefont {W.}~\bibnamefont {Casteels}}\ and\ \bibinfo {author} {\bibfnamefont {C.}~\bibnamefont {Ciuti}},\ }\bibfield  {title} {\bibinfo {title} {Quantum entanglement in the spatial-symmetry-breaking phase transition of a driven-dissipative {{Bose-Hubbard}} dimer},\ }\href {https://doi.org/10.1103/PhysRevA.95.013812} {\bibfield  {journal} {\bibinfo  {journal} {Phys. Rev. A}\ }\textbf {\bibinfo {volume} {95}},\ \bibinfo {pages} {013812} (\bibinfo {year} {2017})}\BibitemShut {NoStop}%
\bibitem [{\citenamefont {Delmonte}\ \emph {et~al.}(2025)\citenamefont {Delmonte}, \citenamefont {Li}, \citenamefont {Passarelli}, \citenamefont {Song}, \citenamefont {Barberena}, \citenamefont {Rey},\ and\ \citenamefont {Fazio}}]{delmonteMeasurementinducedPhaseTransitions2025}%
  \BibitemOpen
  \bibfield  {author} {\bibinfo {author} {\bibfnamefont {A.}~\bibnamefont {Delmonte}}, \bibinfo {author} {\bibfnamefont {Z.}~\bibnamefont {Li}}, \bibinfo {author} {\bibfnamefont {G.}~\bibnamefont {Passarelli}}, \bibinfo {author} {\bibfnamefont {E.~Y.}\ \bibnamefont {Song}}, \bibinfo {author} {\bibfnamefont {D.}~\bibnamefont {Barberena}}, \bibinfo {author} {\bibfnamefont {A.~M.}\ \bibnamefont {Rey}},\ and\ \bibinfo {author} {\bibfnamefont {R.}~\bibnamefont {Fazio}},\ }\bibfield  {title} {\bibinfo {title} {Measurement-induced phase transitions in monitored infinite-range interacting systems},\ }\href {https://doi.org/10.1103/PhysRevResearch.7.023082} {\bibfield  {journal} {\bibinfo  {journal} {Physical Review Research}\ }\textbf {\bibinfo {volume} {7}},\ \bibinfo {pages} {023082} (\bibinfo {year} {2025})}\BibitemShut {NoStop}%
\bibitem [{\citenamefont {Yuan}\ \emph {et~al.}(2019)\citenamefont {Yuan}, \citenamefont {Endo}, \citenamefont {Zhao}, \citenamefont {Li},\ and\ \citenamefont {Benjamin}}]{yuanTheoryVariationalQuantum2019a}%
  \BibitemOpen
  \bibfield  {author} {\bibinfo {author} {\bibfnamefont {X.}~\bibnamefont {Yuan}}, \bibinfo {author} {\bibfnamefont {S.}~\bibnamefont {Endo}}, \bibinfo {author} {\bibfnamefont {Q.}~\bibnamefont {Zhao}}, \bibinfo {author} {\bibfnamefont {Y.}~\bibnamefont {Li}},\ and\ \bibinfo {author} {\bibfnamefont {S.~C.}\ \bibnamefont {Benjamin}},\ }\bibfield  {title} {\bibinfo {title} {Theory of variational quantum simulation},\ }\href {https://doi.org/10.22331/q-2019-10-07-191} {\bibfield  {journal} {\bibinfo  {journal} {Quantum}\ }\textbf {\bibinfo {volume} {3}},\ \bibinfo {pages} {191} (\bibinfo {year} {2019})}\BibitemShut {NoStop}%
\bibitem [{\citenamefont {Kloeden}\ \emph {et~al.}(1992)\citenamefont {Kloeden}, \citenamefont {Platen}, \citenamefont {Kloeden},\ and\ \citenamefont {Platen}}]{kloeden1992stochastic}%
  \BibitemOpen
  \bibfield  {author} {\bibinfo {author} {\bibfnamefont {P.~E.}\ \bibnamefont {Kloeden}}, \bibinfo {author} {\bibfnamefont {E.}~\bibnamefont {Platen}}, \bibinfo {author} {\bibfnamefont {P.~E.}\ \bibnamefont {Kloeden}},\ and\ \bibinfo {author} {\bibfnamefont {E.}~\bibnamefont {Platen}},\ }\href@noop {} {\emph {\bibinfo {title} {Stochastic differential equations}}}\ (\bibinfo  {publisher} {Springer},\ \bibinfo {year} {1992})\BibitemShut {NoStop}%
\bibitem [{Note1()}]{Note1}%
  \BibitemOpen
  \bibinfo {note} {With real variational parameters, the tangent space $T_{\protect \bm {\theta }}\protect \mathcal {M}$ is regarded as a real vector space endowed with the inner product $(\psi _1,\psi _2)=\Re \braket {\psi _1}{\psi _2}$. The corresponding orthogonal projection is then equivalent to the McLachlan variational principle~\cite {yuanTheoryVariationalQuantum2019a}, which minimizes the Hilbert-space distance between the exact and variational increments.}\BibitemShut {Stop}%
\bibitem [{\citenamefont {Le~Boit{\'e}}\ \emph {et~al.}(2013)\citenamefont {Le~Boit{\'e}}, \citenamefont {Orso},\ and\ \citenamefont {Ciuti}}]{leboiteSteadyStatePhasesTunnelingInduced2013}%
  \BibitemOpen
  \bibfield  {author} {\bibinfo {author} {\bibfnamefont {A.}~\bibnamefont {Le~Boit{\'e}}}, \bibinfo {author} {\bibfnamefont {G.}~\bibnamefont {Orso}},\ and\ \bibinfo {author} {\bibfnamefont {C.}~\bibnamefont {Ciuti}},\ }\bibfield  {title} {\bibinfo {title} {Steady-{{State Phases}} and {{Tunneling-Induced Instabilities}} in the {{Driven Dissipative Bose-Hubbard Model}}},\ }\href {https://doi.org/10.1103/PhysRevLett.110.233601} {\bibfield  {journal} {\bibinfo  {journal} {Physical Review Letters}\ }\textbf {\bibinfo {volume} {110}},\ \bibinfo {pages} {233601} (\bibinfo {year} {2013})}\BibitemShut {NoStop}%
\bibitem [{Note2()}]{Note2}%
  \BibitemOpen
  \bibinfo {note} {In the frame rotating at the drive frequency with $\hbar =1$.}\BibitemShut {Stop}%
\bibitem [{\citenamefont {Worth}\ and\ \citenamefont {Burghardt}(2003)}]{worthFullQuantumMechanical2003}%
  \BibitemOpen
  \bibfield  {author} {\bibinfo {author} {\bibfnamefont {G.~A.}\ \bibnamefont {Worth}}\ and\ \bibinfo {author} {\bibfnamefont {I.}~\bibnamefont {Burghardt}},\ }\bibfield  {title} {\bibinfo {title} {Full quantum mechanical molecular dynamics using {{Gaussian}} wavepackets},\ }\href {https://doi.org/10.1016/S0009-2614(02)01920-6} {\bibfield  {journal} {\bibinfo  {journal} {Chemical Physics Letters}\ }\textbf {\bibinfo {volume} {368}},\ \bibinfo {pages} {502} (\bibinfo {year} {2003})}\BibitemShut {NoStop}%
\bibitem [{\citenamefont {Richings}\ \emph {et~al.}(2015)\citenamefont {Richings}, \citenamefont {Polyak}, \citenamefont {Spinlove}, \citenamefont {Worth}, \citenamefont {Burghardt},\ and\ \citenamefont {Lasorne}}]{richingsQuantumDynamicsSimulations2015}%
  \BibitemOpen
  \bibfield  {author} {\bibinfo {author} {\bibfnamefont {G.}~\bibnamefont {Richings}}, \bibinfo {author} {\bibfnamefont {I.}~\bibnamefont {Polyak}}, \bibinfo {author} {\bibfnamefont {K.}~\bibnamefont {Spinlove}}, \bibinfo {author} {\bibfnamefont {G.}~\bibnamefont {Worth}}, \bibinfo {author} {\bibfnamefont {I.}~\bibnamefont {Burghardt}},\ and\ \bibinfo {author} {\bibfnamefont {B.}~\bibnamefont {Lasorne}},\ }\bibfield  {title} {\bibinfo {title} {Quantum dynamics simulations using {{Gaussian}} wavepackets: The {{vMCG}} method},\ }\href {https://doi.org/10.1080/0144235X.2015.1051354} {\bibfield  {journal} {\bibinfo  {journal} {International Reviews in Physical Chemistry}\ }\textbf {\bibinfo {volume} {34}},\ \bibinfo {pages} {269} (\bibinfo {year} {2015})}\BibitemShut {NoStop}%
\bibitem [{\citenamefont {Lasser}\ and\ \citenamefont {Lubich}(2020)}]{lasserComputingQuantumDynamics2020}%
  \BibitemOpen
  \bibfield  {author} {\bibinfo {author} {\bibfnamefont {C.}~\bibnamefont {Lasser}}\ and\ \bibinfo {author} {\bibfnamefont {C.}~\bibnamefont {Lubich}},\ }\bibfield  {title} {\bibinfo {title} {Computing quantum dynamics in the semiclassical regime},\ }\href {https://doi.org/10.1017/S0962492920000033} {\bibfield  {journal} {\bibinfo  {journal} {Acta Numerica}\ }\textbf {\bibinfo {volume} {29}},\ \bibinfo {pages} {229} (\bibinfo {year} {2020})},\ \Eprint {https://arxiv.org/abs/2002.00624} {arXiv:2002.00624} \BibitemShut {NoStop}%
\bibitem [{Note3()}]{Note3}%
  \BibitemOpen
  \bibinfo {note} {For a complex parametrization, we split the real and imaginary parts and work with the effective real parameters.}\BibitemShut {Stop}%
\bibitem [{Note4()}]{Note4}%
  \BibitemOpen
  \bibinfo {note} {Due to the overcompleteness of the Gaussian basis, we employ a standard Tikhonov regularization for numerical stability, which amounts to adding a small diagonal shift $\epsilon $ (typically $\in [10^{-12},10^{-6}]$) to the quantum geometric tensor $S$ before inverting it.}\BibitemShut {Stop}%
\bibitem [{\citenamefont {R{\"u}emelin}(1982)}]{ruemelinNumericalTreatmentStochastic1982}%
  \BibitemOpen
  \bibfield  {author} {\bibinfo {author} {\bibfnamefont {W.}~\bibnamefont {R{\"u}emelin}},\ }\bibfield  {title} {\bibinfo {title} {Numerical {{Treatment}} of {{Stochastic Differential Equations}}},\ }\href {https://doi.org/10.1137/0719041} {\bibfield  {journal} {\bibinfo  {journal} {SIAM Journal on Numerical Analysis}\ }\textbf {\bibinfo {volume} {19}},\ \bibinfo {pages} {604} (\bibinfo {year} {1982})}\BibitemShut {NoStop}%
\bibitem [{Note5()}]{Note5}%
  \BibitemOpen
  \bibinfo {note} {Due to the spatial symmetry of the Liouvillian, the (unique) steady state will have uniform population even in the mean-field symmetry-broken regime~\cite {mingantiSpectralTheoryLiouvillians2018a}. Instead, $\langle {\protect \hat {n}}_\pi \rangle $ can remain nonzero both on the Lindblad level and along trajectories.}\BibitemShut {Stop}%
\bibitem [{\citenamefont {Casteels}\ \emph {et~al.}(2017)\citenamefont {Casteels}, \citenamefont {Fazio},\ and\ \citenamefont {Ciuti}}]{casteelsCriticalDynamicalProperties2017}%
  \BibitemOpen
  \bibfield  {author} {\bibinfo {author} {\bibfnamefont {W.}~\bibnamefont {Casteels}}, \bibinfo {author} {\bibfnamefont {R.}~\bibnamefont {Fazio}},\ and\ \bibinfo {author} {\bibfnamefont {C.}~\bibnamefont {Ciuti}},\ }\bibfield  {title} {\bibinfo {title} {Critical dynamical properties of a first-order dissipative phase transition},\ }\href {https://doi.org/10.1103/PhysRevA.95.012128} {\bibfield  {journal} {\bibinfo  {journal} {Phys. Rev. A}\ }\textbf {\bibinfo {volume} {95}},\ \bibinfo {pages} {012128} (\bibinfo {year} {2017})}\BibitemShut {NoStop}%
\bibitem [{Note6()}]{Note6}%
  \BibitemOpen
  \bibinfo {note} {The ``mean-field'' trajectory entanglement is obtained via a semiclassical theory recently developed by some of us as detailed in~\cite {liEmergentDeterministicEntanglement2025}.}\BibitemShut {Stop}%
\bibitem [{\citenamefont {Li}\ \emph {et~al.}(2025{\natexlab{a}})\citenamefont {Li}, \citenamefont {Delmonte},\ and\ \citenamefont {Fazio}}]{liEmergentDeterministicEntanglement2025}%
  \BibitemOpen
  \bibfield  {author} {\bibinfo {author} {\bibfnamefont {Z.}~\bibnamefont {Li}}, \bibinfo {author} {\bibfnamefont {A.}~\bibnamefont {Delmonte}},\ and\ \bibinfo {author} {\bibfnamefont {R.}~\bibnamefont {Fazio}},\ }\bibfield  {title} {\bibinfo {title} {Emergent deterministic entanglement dynamics in monitored infinite-range bosonic systems},\ }\href {https://doi.org/10.1103/bhst-127b} {\bibfield  {journal} {\bibinfo  {journal} {Physical Review B}\ }\textbf {\bibinfo {volume} {112}},\ \bibinfo {pages} {104315} (\bibinfo {year} {2025}{\natexlab{a}})}\BibitemShut {NoStop}%
\bibitem [{\citenamefont {Li}\ \emph {et~al.}(2022)\citenamefont {Li}, \citenamefont {Claude}, \citenamefont {Boulier}, \citenamefont {Giacobino}, \citenamefont {Glorieux}, \citenamefont {Bramati},\ and\ \citenamefont {Ciuti}}]{Zejian}%
  \BibitemOpen
  \bibfield  {author} {\bibinfo {author} {\bibfnamefont {Z.}~\bibnamefont {Li}}, \bibinfo {author} {\bibfnamefont {F.}~\bibnamefont {Claude}}, \bibinfo {author} {\bibfnamefont {T.}~\bibnamefont {Boulier}}, \bibinfo {author} {\bibfnamefont {E.}~\bibnamefont {Giacobino}}, \bibinfo {author} {\bibfnamefont {Q.}~\bibnamefont {Glorieux}}, \bibinfo {author} {\bibfnamefont {A.}~\bibnamefont {Bramati}},\ and\ \bibinfo {author} {\bibfnamefont {C.}~\bibnamefont {Ciuti}},\ }\bibfield  {title} {\bibinfo {title} {Dissipative phase transition with driving-controlled spatial dimension and diffusive boundary conditions},\ }\href {https://doi.org/10.1103/PhysRevLett.128.093601} {\bibfield  {journal} {\bibinfo  {journal} {Phys. Rev. Lett.}\ }\textbf {\bibinfo {volume} {128}},\ \bibinfo {pages} {093601} (\bibinfo {year} {2022})}\BibitemShut {NoStop}%
\bibitem [{\citenamefont {Li}\ \emph {et~al.}(2025{\natexlab{b}})\citenamefont {Li}, \citenamefont {Delmonte}, \citenamefont {Turkeshi},\ and\ \citenamefont {Fazio}}]{liMonitoredLongrangeInteracting2025a}%
  \BibitemOpen
  \bibfield  {author} {\bibinfo {author} {\bibfnamefont {Z.}~\bibnamefont {Li}}, \bibinfo {author} {\bibfnamefont {A.}~\bibnamefont {Delmonte}}, \bibinfo {author} {\bibfnamefont {X.}~\bibnamefont {Turkeshi}},\ and\ \bibinfo {author} {\bibfnamefont {R.}~\bibnamefont {Fazio}},\ }\bibfield  {title} {\bibinfo {title} {Monitored long-range interacting systems: Spin-wave theory for quantum trajectories},\ }\href {https://doi.org/10.1038/s41467-025-59557-w} {\bibfield  {journal} {\bibinfo  {journal} {Nature Communications}\ }\textbf {\bibinfo {volume} {16}},\ \bibinfo {pages} {4329} (\bibinfo {year} {2025}{\natexlab{b}})}\BibitemShut {NoStop}%
\bibitem [{\citenamefont {Li}\ \emph {et~al.}(2026)\citenamefont {Li}, \citenamefont {Delmonte},\ and\ \citenamefont {Fazio}}]{liGeneralizedStochasticSpinwave2026}%
  \BibitemOpen
  \bibfield  {author} {\bibinfo {author} {\bibfnamefont {Z.}~\bibnamefont {Li}}, \bibinfo {author} {\bibfnamefont {A.}~\bibnamefont {Delmonte}},\ and\ \bibinfo {author} {\bibfnamefont {R.}~\bibnamefont {Fazio}},\ }\bibfield  {title} {\bibinfo {title} {Generalized stochastic spin-wave theory for open quantum spin systems},\ }\href {https://doi.org/10.1103/d942-3lmt} {\bibfield  {journal} {\bibinfo  {journal} {Physical Review B}\ }\textbf {\bibinfo {volume} {113}},\ \bibinfo {pages} {214324} (\bibinfo {year} {2026})}\BibitemShut {NoStop}%
\bibitem [{\citenamefont {Minganti}\ \emph {et~al.}(2018)\citenamefont {Minganti}, \citenamefont {Biella}, \citenamefont {Bartolo},\ and\ \citenamefont {Ciuti}}]{mingantiSpectralTheoryLiouvillians2018a}%
  \BibitemOpen
  \bibfield  {author} {\bibinfo {author} {\bibfnamefont {F.}~\bibnamefont {Minganti}}, \bibinfo {author} {\bibfnamefont {A.}~\bibnamefont {Biella}}, \bibinfo {author} {\bibfnamefont {N.}~\bibnamefont {Bartolo}},\ and\ \bibinfo {author} {\bibfnamefont {C.}~\bibnamefont {Ciuti}},\ }\bibfield  {title} {\bibinfo {title} {Spectral theory of {{Liouvillians}} for dissipative phase transitions},\ }\href {https://doi.org/10.1103/PhysRevA.98.042118} {\bibfield  {journal} {\bibinfo  {journal} {Physical Review A}\ }\textbf {\bibinfo {volume} {98}},\ \bibinfo {pages} {042118} (\bibinfo {year} {2018})}\BibitemShut {NoStop}%
\bibitem [{\citenamefont {Gardiner}\ and\ \citenamefont {Zoller}(2004)}]{gardiner2004quantumnoise}%
  \BibitemOpen
  \bibfield  {author} {\bibinfo {author} {\bibfnamefont {C.}~\bibnamefont {Gardiner}}\ and\ \bibinfo {author} {\bibfnamefont {P.}~\bibnamefont {Zoller}},\ }\href@noop {} {\emph {\bibinfo {title} {Quantum Noise: A Handbook of Markovian and Non-Markovian Quantum Stochastic Methods with Applications to Quantum Optics}}},\ Springer Series in Synergetics\ (\bibinfo  {publisher} {Springer},\ \bibinfo {year} {2004})\BibitemShut {NoStop}%
\end{thebibliography}%

\appendix

\onecolumngrid

\begin{center}
\large\bfseries
    End Matter
\end{center}

\setcounter{equation}{0}
\setcounter{figure}{0}
\setcounter{table}{0}
\makeatletter
\renewcommand{\theequation}{A\arabic{equation}}
\renewcommand{\thefigure}{A\arabic{figure}}
\twocolumngrid

\section{Appendix A: Stratonovich form of the heterodyne stochastic
\schr{} equation}
\label{app:ito-strat}
In this Appendix we derive the Stratonovich SSE. We refer the reader to Refs.~\cite{kloeden1992stochastic,gardiner2004quantumnoise} for comprehensive treatments, and to Ref.~\cite{zoller1997quantumnoise} for a pedagogical discussion in the context of stochastic \schr{} equations.
For a stochastic process $X_t$ obeying the It\^{o} stochastic differential equation (SDE)
\begin{equation}
    \label{eq:sde-generic}
    \d X_t = a(X_t)\,\dt + b(X_t)\,\d W_t\,,
\end{equation}
where $W_t$ is a real Wiener process satisfying $\d W^2=\d t$, its conversion to the Stratonovich form is done via
\begin{equation}
    \d X_t = \left[ a(X_t) - \frac{1}{2} b(X_t)\, b'(X_t) \right]\dt + b(X_t)\circ\d W_t\,,
\end{equation}
which is an equivalent representation of the \emph{same} stochastic process. 
In practice, the conversion is most easily carried out with the operational rule
\begin{equation}
    \label{eq:strat-rule}
    b(X_t)\circ\d W_t = b(X_t)\,\d W_t + \frac{1}{2}\,\d b(X_t)\,\d W_t\,,
\end{equation}
where the $\d b$ term is expanded with It\^{o}'s lemma and the only surviving contribution is from the finite quadratic variance $\d W^2=\d t$, resulting in a correction to the drift part. Equation~\eqref{eq:strat-rule} generalizes verbatim to the complex noise considered in the main text, which can be decomposed into two independent Wiener processes as $\d Z=(\d W_x+\rmi \d W_y)/\sqrt{2}$, and the multi-dimensional generalization is also straightforward, by applying the rule~\eqref{eq:strat-rule} to each noise channel independently.
For the It\^{o} SSE~\eqref{eq:sse-ito}, we therefore have,
\begin{equation}
    \label{eq:strat-term}
    \hat{M}_m\ket{\psi}\circ\d Z^*_{m} = \hat{M}_m\ket{\psi}\,\d Z^*_{m} + \frac{1}{2}\,\d(\hat{M}_m\ket{\psi})\d Z^*_{m}\,,
\end{equation}
and the It\^{o} calculus gives the expansion,
\begin{equation}
    \d(\hat{M}_m\ket{\psi}) = (\d\hat{M}_m)\ket{\psi} + \hat{M}_m\,\d\ket{\psi} + (\d\hat{M}_m)\,\d\ket{\psi}\,,
\end{equation}
where we note that both $\hat{M}_m$ are $\ket{\psi}$ are stochastic (due to the state-dependence in $\hat{M}_m$) and should be further expanded using the (It\^{o}) differential product rule. The final expression for the correction term takes the following simple form,
\begin{equation}
    \frac{1}{2}\,(\d\hat{M}_m)\,\d Z^*_{m} = -\frac{1}{2}\left( \langle\dLhat_m\Lhat_m\rangle - \lvert\langle\Lhat_m\rangle\rvert^2 \right)\dt\,.
\end{equation}
which gives the Stratonovich drift term in Eq.~\eqref{eq:drift-strat}.

\section{Appendix B: Analytical assembly of the sTDVP equations}
The multi-Gaussian ansatz enables the fully analytical evaluation of the quantum geometric tensor $S$ and the force vectors $\Vvec$, $\Yvec^{(m)}$ in the sTDVP equation~\eqref{eq:stdvp-eom}.
We define a generating function for an arbitrary operator $\Ohat$, as the matrix element between two states with independent variational parameters,
\bea\label{eq:generating-function}
    \zcal^{[\Ohat]}(\thetavec^L,\thetavec^R)&=\bra{\psi[\thetavec^L]}\Ohat\ket{\psi[\thetavec^R]}\\&=\sum_{k,k'=1}^{N_c}\bra{\gcal[\thetavec_k^L]}\Ohat\ket{\gcal[\thetavec_{k'}^R]}\,,
\eea
with the Gaussian state $\ket{\gcal[\thetavec_k]}$ defined such that $\braket{\xvec}{\gcal[\thetavec_k]}=\gcal(\xvec;\thetavec_k)$ [cf. Eq.~\eqref{eq:Gkx}]. The sTDVP ingredients can then be expressed as $\thetavec-$derivatives of this generating function. For the quantum geometric tensor $S_{ij}=\braket{v_i}{v_j}$, the relevant operator is simply $\Ohat=\idhat$,
\bea\label{eq:qgt-generating-function}
    S_{ij}[\thetavec]=\partial_{\theta^L_i}\partial_{\theta^R_j}\zcal^{[\idhat]}(\thetavec^L,\thetavec^R)|_{\thetavec^L=\thetavec^R=\thetavec}\,.
\eea
Similarly, the drift forces  $V_{j} = \bra{v_j}\hat{D}_{\mathrm{S}}\ket{\psi}$, $Y^{(m)}_{j} = \bra{v_j}\hat{M}_m\ket{\psi}$ correspond to first-order derivatives involving the drift and noise operators,
\bea\label{eq:force-generating-function}
    V_j(\thetavec)&=\partial_{\theta^L_j}\zcal^{[\Dhat]}(\thetavec^L,\thetavec^R)|_{\thetavec^L=\thetavec^R=\thetavec}\,,\\
    Y_j^{(m)}(\thetavec)&=\partial_{\theta^L_j}\zcal^{[\Mhat_m]}(\thetavec^L,\thetavec^R)|_{\thetavec^L=\thetavec^R=\thetavec}\,.
\eea
We note that the state-dependent expectation values involved in $\Dhat$ and $\Mhat_{m}$, such as $\langle\Lhat_m\rangle$, are evaluated on the ``right'' state $\ket{\psi[\thetavec^R]}$, which are, therefore, not affected by the ``left'' differentiation $\partial_{\thetavec^L}$. In practice, they are evaluated at the current state parameter $\thetavec(t)$ prior to differentiation and treated as fixed scalar coefficients.

As the Bose-Hubbard Hamiltonian and Lindblad operators are finite-degree polynomials in the creation and annihilation operators, i.e., of the form $\Ohat=P(\aaa_i,\daaa_i)$ for some polynomial $P$, the position-space representation is also a polynomial $\pcal(\xvec,\partial_{\xvec})$ via the canonical mapping~\eqref{eq:canonical-mapping}. The generating function then becomes,
\bea
    \zcal^{[\Ohat]}(\thetavec^L,\thetavec^R)&=\sum_{k,k'}\bra{\gcal[\thetavec_k^L]}P(\aaa_i,\daaa_i)\ket{\gcal[\thetavec_{k'}^R]}\\
    &= \sum_{k,k'}\int\d^N \xvec~ \gcal^*(\xvec;\thetavec^L_k)\pcal(\xvec,\partial_{\xvec})\gcal(\xvec;\thetavec^R_{k'})\\
    &= \sum_{k,k'}\int\d^N \xvec~ \qcal(\xvec) \gcal^*(\xvec;\thetavec^L_k)\gcal(\xvec;\thetavec^R_{k'})\,,
\eea
for some finite degree polynomial $\qcal(\xvec)$, since the application of $\pcal(\xvec,\partial_{\xvec})$ on the Gaussian function $\gcal(\xvec;\thetavec^R_{k'})$ results in a polynomial in $\xvec$ multiplying the same Gaussian function. Furthermore, the product of the two Gaussian functions $\gcal^{LR}_{k,k'}(\xvec)\equiv \gcal^*(\xvec;\thetavec^L_k)\gcal(\xvec;\thetavec^R_{k'})$ is also a Gaussian function, which reads
\bea
    \gcal^{LR}_{k,k'}(\xvec) =&~ C^{L*}_k C^R_{k'}\pi^{-\frac{N}{2}}\lvert A^L_k+A^{L*}_k\rvert^{\frac{1}{4}}\lvert A^R_{k'}+A^{R*}_{k'}\rvert^{\frac{1}{4}}\times\\
    &~\exp\left\{ -\xvec^T B \xvec + \vvec^T\xvec + \Theta \right\}\,,\\
    B=&~A^{L*}_k+A^R_{k'}\,,\\
    \vvec =&~ 2A^{L*}_k\qvec^L_k + 2A^R_{k'} \qvec^R_{k'} - \rmi(\pvec^L_k - \pvec^R_{k'})\,,\\
    \Theta =&~ -(\qvec^L_k)^T A^{L*}_k\qvec^L_k - (\qvec^R_{k'})^T A^R_{k'}\qvec^R_{k'} \\&+ \rmi(\pvec^L_k)^T\qvec^L_k - \rmi(\pvec^R_{k'})^T\qvec^R_{k'}\,.
\eea
This allows us to formally rewrite the generating function as,
\bea
    \zcal^{[\Ohat]}(\thetavec^L,\thetavec^R)=\sum_{k,k'}\int\d^N\xvec~ \qcal(\xvec) \gcal^{LR}_{k,k'}(\xvec)\,,
\eea
which is a sum of classical expectation values of the polynomial $\qcal(\xvec)$ over (unnormalized) multivariate Gaussian measures $\gcal^{LR}_{k,k'}(\xvec)$, that can be evaluated to a analytical expression thanks to the Wick's theorem. Finally, since differentiating the closed-form expression $ \zcal^{[\Ohat]}(\thetavec^L,\thetavec^R)$ yields further analytical expressions of $(\thetavec^L,\thetavec^R)$, the quantum geometric tensor $S_{ij}$ [eq.~\eqref{eq:qgt-generating-function}] and forces $V_j$, $Y^{(m)}_j$ [eq.~\eqref{eq:force-generating-function}] therefore remain fully analytical.

\section{Appendix C: Convergence of the lattice results with $N_c$}
\begin{figure}[H]
    \centering
    \includegraphics[width=\linewidth]{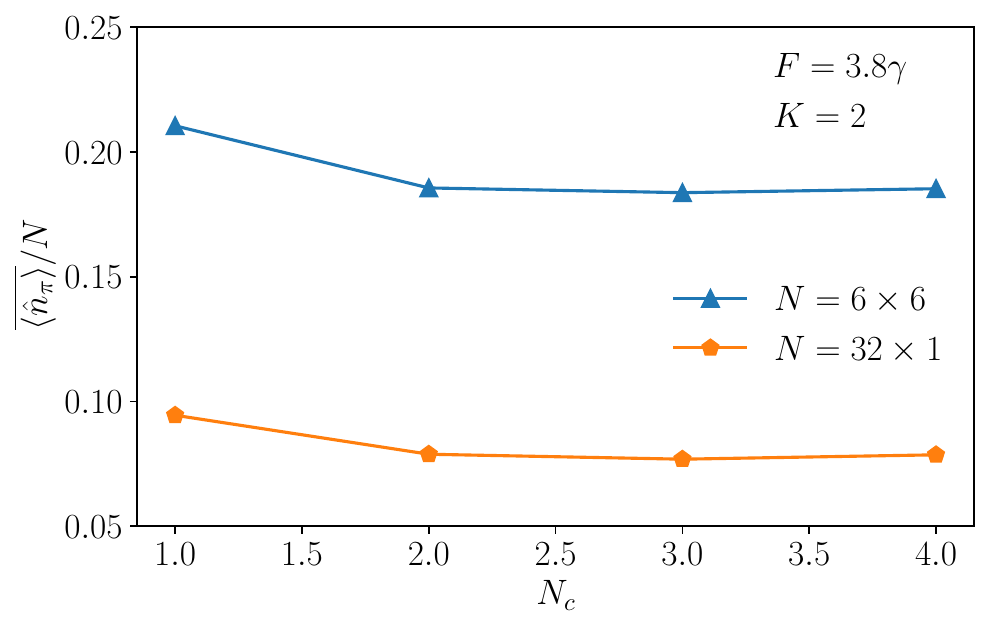}
    \caption{Steady-state normalized $\pi-$mode population obtained with different $N_c$ for two lattice sizes ($32\times 1$ and $6\times 6$) with driving $F=3.8\gamma$ and $K=2$.}
    \label{fig:conv-Nc}
\end{figure}

The convergence of our results on 1D and 2D lattices are verified against the number of Gaussian components $N_c$ [note that for any value of the rank $K$, the multi-Gaussian ansatz~\eqref{eq:psi-vmg} can always be universal approximators with large enough $N_c$]. We present in Fig.~\ref{fig:conv-Nc} the steady-state normalized $\pi-$mode population at $F=3.8\gamma$ (close to the peak of $\langle\nhat_\pi\rangle/N$) obtained with $N_c$ increasing from 1 to 4 (with fixed $K=2$) for the largest lattice sizes in 1D ($32\times 1$) and 2D ($6\times 6$), which are the ``most relevant cases'' supporting our claim of the dimension-dependent phase transition. We observe that for the results reported in the main text, $N_c=4$ and $K=2$ are generally sufficient for achieving convergence in the observable $\langle\nhat_\pi\rangle/N$, while we expect higher values of $N_c$ and $K$ to be necessary for converging in entanglement measures.

\end{document}